\documentstyle[epsf,epsfig,preprint,aps,amssymb]{revtex}
\draft \preprint{SNUTP 02/020}
\begin{document}
\title{\Large\bf $Z_2$ orbifold compactification of heterotic string
and 6D $SO(16)$ and $E_7\times SU(2)$ flavor unification models}
\author{Kang-Sin Choi\footnote{ugha@th.physik.uni-bonn.de} and Jihn E.
Kim\footnote{jekim@th.phyik.uni-bonn.de}}
\address{
Physikalisches Institut, Universit\"{a}t Bonn,
D53115 Bonn, Germany, and \\
School of Physics, Seoul National University, Seoul 151-747,
Korea}\maketitle

\begin{abstract}
A $Z_2$ orbifold compactification of the heterotic string is
considered. The resulting 6D GUT groups can be $SO(16)$ or 
$E_7\times SU(2)$ plus some hidden sector groups.
The $N=4$ supersymmetry is reduced to
$N=2$. In particular, the $SO(16)$ 6D model with one spinor
representation {\bf 128} can reduce to the previous 
5D \  $SO(16)$ or $SO(14)$ family
unification models after compactifying the sixth dimension.
To obtain one spinor, we have
to take into account the left-over center of $SO(16)$.
We also comment on the $E_7\times SU(2)$ model.
\\
\vskip 0.5cm\noindent [Key words: $Z_2$ orbifold, superstring, 6D
SO(16) GUF, $SU(2)$ holonomy]
\end{abstract}


\newcommand{\bea}{\begin{eqnarray}}
\newcommand{\eea}{\end{eqnarray}}
\def\beq{\begin{equation}}
\def\eeq{\end{equation}}

\def\one{\bf 1}
\def\two{\bf 2}
\def\five{\bf 5}
\def\ten{\bf 10}
\def\tenb{\overline{\bf 10}}
\def\fiveb{\overline{\bf 5}}
\def\threeb{{\bf\overline{3}}}
\def\three{{\bf 3}}
\def\fb{{\overline{F}\,}}
\def\hb{{\overline{h}}}
\def\Hb{{\overline{H}\,}}

\def\slash#1{#1\!\!\!\!\!\!/}
\def\hf{\frac12}

\def\A{{\cal A}}
\def\Q{{\cal Q}}

\newcommand{\debug}{\emph{!!! CHECK !!!}}

\newcommand{\dd}{\mathrm{d}\,}
\newcommand{\Tr}{\mathrm{Tr}}
\newcommand{\drep}[2]{(\mathbf{#1},\mathbf{#2})}

\newpage

\section{Introduction}

With the discovery of the top quark, the three family standard
model has been filled and seems to be the theory below 100~GeV
energy scale. Presumably, the observed weak CP violation is the 
most compelling reason for the three family standard model(SM),   
but it lacks the theoretical reasoning of why the same fermion
representation repeats three times. In this regard, the grand  
unification of families or flavors(GUF) attempts to unify the SM
gauge couplings without the repetition of fermionic
representations of the grand unification gauge group\cite{georgi},
which seems to be the most promising rationale toward interpreting
multi generations. In early days of GUF, indeed there appeared
models without repeated fermionic
representations\cite{georgi,kimsu7}.

But with the advent of superstring models, the GUF idea seems to
be automatically implemented in superstring models. In particular,
the $Z_3$ orbifold models usually give 4D $N=1$
supersymmetric(SUSY) three family models and $Z_3$ seemed to be 
one of the fundamental principles of compactification of extra  
internal space\cite{dixon,iknq}. Some orbifold models are very  
close to the standard model\cite{iknq}, but in most cases, there
appear charged SM singlet fermions, which may not be consistent
with the observed weak mixing angle $\sin^2\theta_W|_{\rm 100\   
GeV}\simeq 0.232$. Therefore, one of the key points to require in
the compactification is not allowing unfamiliar fermions(UF). For
example, vectorlike charged fermions are UF's since they are not
appearing in the SM. For this reason, the flipped
$SU(5)$\cite{su51} attracted a great deal of attention in the  
fermionic construction of 4D string models\cite{ellis}. However,
the orbifold compactification of $SU(5)\times U(1)$ seemed to be
not easy. Recently, a 5D $SO(14)$ model without UF's has been
considered to give a 4D three family flipped $SU(5)$ model which
does not violate major low energy observations\cite{barr,hk1}.

From a 10D superstring theory, one has to compactify {\it six 
extra internal spaces to obtain a 4D SUSY theory}\cite{candelas}.
Most widely considered superstring theory is the heterotic
$E_8\times E_8^\prime$. But the field theoretic orbifold
compactification\cite{kawamura} need the string compactification  
of the internal space smaller than six, so that we can consider
SUSY field theory in $D\ge 5$. For the possibility of
complexifying the internal space, we consider compactifying even
dimensions, and hence our consideration of {\it the internal space
is four dimensions, leading to a 6D SUSY field theory model.}   
The field theoretic compactification\cite{kawamura} need the 
string compactification
of the internal space smaller than six, so that we can consider
SUSY field theory in $D\ge 5$. For the possibility of
complexifying the internal space, we consider compactifying even
dimensions, and hence our consideration of {\it the internal space
is four dimensions, leading to a 6D SUSY field theory model.} 

The heterotic string theory has $N=4$ in the 4D viewpoint. To
obtain chiral fermions, we have to reduce $N=4$ down to $N=1$. The
most famous internal space for this purpose is the Calabi-Yau
space with $SU(3)$ holonomy\cite{candelas}. The $Z_3$ orbifold
also reduces $N=4$ down to $N=1$\cite{dixon}. It can be understood
by observing that blowing up the singularities of the orbifold
fixed points leads to the Calabi-Yau space with the $SU(3)$
holonomy, because $Z_3$ is the center of $SU(3)$. To obtain a 6D
model by compactifying four internal space, we cannot have an   
$SU(3)$ holonomy since 6D SUSY theory must be $N=2$. But we find  
that $Z_2$ orbifold works in reducing $N=4$ down to $N=2$. It is
because an $SO(4)$ vector {\bf 4} is $({\bf 2,2})$ under
$SU(2)\times SU(2)$ and giving a vacuum expectation value to {\bf
4} can break one $SU(2)$ but leaves the other $SU(2)$ unbroken,
which leads to an $SU(2)$ holonomy. Then SUSY is reduced by half.
Since $Z_2$ is the center of $SU(2)$, we obtain 6D $N=2$(in 4D
viewpoint) theory. To obtain an $N=1$ theory, one
has to introduce another $Z_2'$ to break the remaining
$SU(2)$, which we do not consider in this paper.  With the $Z_2$
orbifold, we obtain the
simplest 6D orbifold family unification model $SO(16)$, which can 
be compared to the simplest 4D orbifold model $E_6\times
SU(3)$\cite{dixon}. It can lead to 5D flavor unification
models considered in Refs.\cite{barr,hk1}.
Also, it is possible to consider a 6D $E_7\times SU(2)$ model
as a flavor unification model.

\section{$Z_2$ orbifold}

In this spirit, let us proceed to consider the orbifold
compactification of $T^m/{\bf Z}_n$ with $m=4$ and $n=2$ in
mind.\footnote{Even though we are interested in $m=4$ and $n=2$,
some formulae include $m$ and $n$ explicitly to compare with the
well-known case $m=6$ and $n=3$\cite{dixon}.} The twisting of the 
$m$-dimensional internal space and gauge groups are,

\begin{eqnarray}\label{twist}
 \mbox{internal space}:&\exp(2\pi i[\frac12 J_{67}+\frac12 J_{89}]),
 \nonumber\\
 \mbox{gauge groups}:&v_{1I} =\frac1n(w_I), \quad (E_8: I=1,\dots,8) \\
                     &v_{2I} =\frac1n(w_I^\prime), \quad
(E_8^\prime:I=9,\dots,16). \nonumber
\end{eqnarray}

We define $\alpha$ is the eigenvalue of the internal space
rotation,
$$
\exp \left(2\pi
i\sum_{a=1}^{m/2}\frac{r_a}{n}J_{2a+4,2a+5}\right),
$$
satisfying $\alpha^n=1$.

There are conditions to be satisfied\cite{dixon}. 

$$ \sum r_a = \sum w_I = \sum w_I' = 0 \mbox \quad \mbox{mod 2.} $$

\begin{equation}\label{cond2}
\sum r_a^2 = \sum w_I^2 + \sum {w'_I}^2 \quad \mbox{mod
}\epsilon_nn, 
\end{equation}
where $\epsilon_N=2(1)$ for even(odd) $n$. The first equation in 
Eq.~(\ref{cond2}) is from the definition 
of twist of order $n$ up to fermionic degree of freedom,
and the second one is the modular invariance condition.

There are only few possibilities for the lattice vector $v_1$
and $v_2$ of $E_8\times E_8'$ satisfying these conditions. For $2
v_i$ should be again lattice vector in $E_8$, $v_i^2$ is $\frac12$
times an integer. Also it is known that any lattice vector in
$E_8$ is within the distance 1 from some lattice point, so $v_i^2
\le 1$. There are only three allowed combinations which are shown
in Table I.\vskip 0.5cm

\begin{center} Table I. {\it $Z_2$ orbifolds of the 
heterotic string.} \vskip
0.3cm
\begin{tabular}{|c|ccccc|}\hline   
Case & $E_8$ shift &\qquad& $E_8'$ shift &\qquad& 6D gauge
group \\
\hline 
 (i) &
$(\frac12~ \frac12~0~0~0~0~0~0 )$ &\qquad& $(0~0~0~0~0~0~0~0)$  &\qquad& 
$E_7 \times SU(2) \times E_8'
$ \\
(ii) & $(1 ~0 ~0~ 0 ~0~0~0~0)$ &\quad&
$(\hf~\hf~ 0~0~0~0~0~0)$ &\quad& $ SO(16)
\times SU(2)'\times E_7' $ \\
(iii) & $(\frac12 ~\frac12 ~\frac12~ \frac12 ~0~0~0~0)$ &\quad&
$(\frac12~\frac12~ 0~0~0~0~0~0)$ &\quad& $ SO(16)
\times SU(2)' \times E_7' $ \\
\hline
\end{tabular}\\
\end{center}
\vskip 0.5cm
Case (i) breaks down to $E_7\times SU(2)\times
E_8^\prime$. Cases (ii) and (iii) break down to
$SO(16)\times  SU(2)'\times E_7'$. The cases $(v_1^2=1,v_2^2=0),
(v_1^2=v_2^2=1)$ with entries $\frac12$, 
and $(v_1^2 =v_2^2=\frac12)$ do not satisfy the
two conditions given above. We will mainly discuss $SO(16)$ and comment
on $E_7\times SU(2)$ flavor unification group at the end.

\subsection{$SO(16)$}

Let us consider the $SO(16)$ first which can give 5D
flavor unification models\cite{barr,hk1}.
Even though Cases (ii) and (iii) give the same theory, we will treat
them separately, since the introduction of Wilson lines can be studied
more easily with two different shift vectors.
It is easy to understand Case (ii). It is simply separating
{\bf 248} of $E_8$ into vector and spinor parts, where the
vector forming the adjoint representation of $SO(16)$ and the spinor
formining the matter fields {\bf 128} of $SO(16)$\cite{dixon}.
Thus, we discuss Case (iii) only in detail since it
has not been discussed in the literature
and can be a potential GUF(grand unification
of families) {$SO(16)$}\cite{barr,hk1}. The gauge group from
$SU(2)^\prime\times E_7'$ can be considered as the ``hidden''
sector needed for SUSY breaking.

Now, from the mass shell condition
\begin{equation}
 {m^2 \over 4} = {p^2 \over 2} - 1
\end{equation}
we have massless states as follows:

On the left-moving side there are

\begin{eqnarray}
  \alpha^0 : & \alpha_{-1}^\mu |0 \rangle \quad \mbox{and {\bf
112}}\\
  \alpha^1 : & \alpha_{-1}^i |0 \rangle, \alpha_{-1}^{\bar i} |0 \rangle
\quad  \mbox{and {\bf 128}}
\end{eqnarray}
where we have not displayed the hidden sector. The corresponding
lattice vectors are shown in the following tables.

\begin{center}
Table II. {\it Root vectors $p_I$ in untwisted sector transforming like
$\alpha^0$. The underlined entries allow permutations and
those in the [] bracket allow even numbers of sign flips.}
\vskip 0.3cm
\begin{tabular}{|cc|}
\hline
vector & number of states \\
\hline
$ (0~0~0~0~\underline{\pm 1~\pm 1~0~0}) $ & 24 \\
$ (\underline{\pm 1~ \pm 1~0~0}~0~0~0~0) $ & 24 \\
$ ([\hf~\hf~\hf~\hf]~[\hf~\hf~\hf~\hf]) $ & 64 \\
\hline
\end{tabular}
\\
\end{center}
In the Tables, we have the convention that the underlined entries
allow premutations and those in the square braket []
allow even numbers of sign flips. We have 112 winding states and from
the oscillators we have 8 $U(1)$ generators. They will form
the adjoint representation {\bf 120} of $SO(16)$.

\begin{center}
Table III. {\it Root vectors $p_I$ in untwisted sector transforming
like $\alpha^1$.}
\vskip 0.3cm  
\begin{tabular}{|cc|}
\hline
vector & number of states \\
\hline
$ (\underline{\pm 1~0~0~0}~\underline{\pm 1~0~0~0}) $ & 64 \\
$ ([-\hf~\hf~\hf~\hf]~[-\hf~\hf\hf\hf]) $ & 64 \\
\hline
\end{tabular}
\end{center}

On the right moving side, there are vectors transforming like
$\alpha^0=1$ and $\alpha^1=e^{2\pi i(1/2)}=-1$,
\begin{eqnarray}
 \alpha^0 : & \tilde b_{-\frac12}^\mu |0 \rangle_{\rm NS},~ |+; {\bf 1}
\rangle_{\rm R} \\
 \alpha^1 : & \tilde b_{-\frac12}^i |0 \rangle_{\rm NS},~
\tilde b_{-\frac12}^{\bar i} |0 \rangle_{\rm NS},~ |-; {\bf 2} 
\rangle_{\rm 
R} 
\end{eqnarray}
where $\tilde b^M_{-\frac12}$s are NS creation   
operators. 
Note that, the first entry of spinorial(Ramond) representation 
is chirality in six dimension. The $|+;{\bf 1} \rangle_R$ (leading 
to gaugino after combining with the left movers) and  
$|-;{\bf 2}\rangle_R$
(leading to the chiral matter after combining with the left movers)  
have opposite chiralities. For $D=2n$, a spinor of $SO(2n)$ can be 
represented by $n$-tuples of spin $\hf$ eigenstates as,
$|s_1~ s_2~ \dots ~\ s_n \rangle$ where $s_i$ can be either 
$\pm \hf$. Then, the chirality is defined by the eigenvalue of 
\begin{equation}\label{chirality}
\Gamma = 2^n s_1 s_2 \dots s_n .
\end{equation}
Compactification of some internal space of even dimensions(e.g., 
from 10D to 6D) kicks out some factors of $s_i$ in 
Eq.~(\ref{chirality}) and the product of the remaining 
$s_i$'s in Eq.~(\ref{chirality}) determines the chirality in the 
uncompactified space. For example, the chirality in 6D in 
our compactification of 10D down to 6D is $\Gamma^7=4 s_1 s_2$.
The ten dimensional {\bf 8} of right-moving ground state is 
decomposed into two {\bf 2}'s and four {\bf 1}'s with {\em opposite} 
chirality in six dimension.

Combining the right movers of the preceeding
paragraph with the left movers into $Z_2$ invariant states, 
we have an adjoint representation {\bf 120} for the gauge multiplet
and a spinor representation {\bf 128} for the hyper-multiplet.
Defining the chirality of the gauge multiplet as left-handed, the
chirality of the hyper-multiplet becomes right-handed. In addition,
there appear the 6D supergravity multiplets, etc.

Notice, however, that combining the right-handed $|{\bf 2} \rangle$ with
the left-handed {\bf 128} seems to give 2 spinors.\footnote{In $Z_3$
orbifold, this is the reason that we obtain three copies of chiral
fermions from the untwisted sector.} But note that the center of
$SO(2n)$ is $Z_2 \times Z_2'$ for an even $n$
and $Z_4$ for an odd $n$\cite{bacry}.
Our orbifolding under $Z_2$ and assigning
at the center of $SO(16)$ still allow $Z_2'$ freedom. Thus, the two
spinors of $SO(16)$ are connected by $SO(16)$ gauge transformation,
and we have to divide the number of spinors by the left-over
center $Z_2'$. Thus, the untwisted sector allows only
one spinor {\bf 128} of $SO(16)$. It is similar to the mechanism
of calculating the domain wall number in axionic models\cite{shafi}.

\subsection{Twisted sector}

In the twisted sector, the twisted mode expansion gives a
different zero point energy. The zero point energy of a bosonic
string 
is given by $\sum_{n=1}^\infty (n + \eta) = -\frac1{24} + 
\frac14 
\eta (1-\eta)
$,
where $\eta=\frac1n$ is a shift.
In our $Z_2$ model, the zero point 
energy is
$  E_0= -1 + 4 \frac 14 \frac 12 \frac12 =- \frac34 $
Then the mass shell condition becomes
\begin{equation}
{m_{\rm R}^2 \over 8} = {m_{\rm L}^2 \over 8} = {(p^I + v^I)^2
\over 2} + \tilde N - \frac34 = 0. \label{twistsct}
\end{equation}
where $\tilde N$ is the oscillator number. For $\tilde N=
\frac{1}{2}$, there is no vector in the $E_8$ lattice, satisfying
Eq.~(\ref{twistsct}).
For $\tilde N=0$, we have some lattice vectors satisfying the above
condition. We can find two sixteen states satisfying Eq.~(\ref{twistsct}) 
given in Table IV.

\begin{center}
Table IV. {\it Root vectors $p_I$ in the twisted sector}
\vskip 0.3cm
\begin{tabular}{|cc|}
\hline
vector & number of states \\
\hline
$ (-\hf~-\hf~-\hf~-\hf~[\hf~\hf\hf\hf])(0~0~0~0~0~0~0~0) $ & 8 \\
$ (\underline{-1~-1~0~0}~0~0~0~0)(0~0~0~0~0~0~0~0) $ & 6 \\
$ (-1~-1~-1~-1~0~0~0~0)(0~0~0~0~0~0~0~0) $ & 1 \\
$ (0~0~0~0~0~0~0~0)(0~0~0~0~0~0~~0) $ & 1 \\
\hline
$ (-\hf~-\hf~-\hf~-\hf~[\hf~\hf\hf\hf])(-1~-1~0~0~0~0~0~0) $ & 8 \\
$ (\underline{-1~-1~0~0}~0~0~0~0)(-1~-1~0~0~0~0~0~0) $ & 6 \\
$ (-1~-1~-1~-1~0~0~0~0)(-1~-1~0~0~0~0~0~0) $ & 1 \\
$ (0~0~0~0~0~0~0~0)(-1~-1~0~0~0~0~~0) $ & 1 \\
\hline
\end{tabular}
\end{center}

Combining with the right-moving states, we have two {\bf 16}'s in the 
twisted sector.

In the $Z_3$ orbifold model,
we encounter only fixed points.
However 
in a general $Z_n$ orbifold model with $n$ even, 
there are also fixed tori 
which  have a different topology and can give a different number
for the twisted sector states. 

For the case of the $Z_2$ orbifold, there are 16 fixed points. However,
the effective multiplicity is 8 rather than 16. 
\footnote{This can be more transparent by reading off 
the partition function. 
In terms of the notation in Ref.\cite{imnq}, 
projector operator is $$P_\theta = 
\frac12 ( \chi_{(\theta,1)} \Delta_\theta^ 0 + \chi_{(\theta,\theta)} 
\Delta 
_\theta^1 ) $$ 
with $\chi_{(\theta^m,\theta^n)}$ is the Euler number of compact 
manifold twisted by $\theta^m, \theta^n$ in the 
each worldsheet direction. Here, 
$\Delta_\theta = \exp 2 \pi i [(p+\tilde v)\tilde v-(r+w+v)v]$ 
where $p$ and $(r+w)$ are the 
lattice of the left and right movers, respectively. In
the $Z_2$ case,  
$\chi_{(\theta,\theta^n)}=16$ where $n=0,1$ and $\Delta_\theta$ 
can be either 1 or $- 1$.}

Therefore, there appear sixteen($8\times 2$)
copies of {\bf 16} from the twisted sector. These have the same
chirality as the matter representation from the untwisted 
sector.\footnote{The $r+\omega$ vector for the right movers 
in the notation of Ref.\cite{imnq} can be $(\hf,\hf,\hf,\hf),
(-\hf,-\hf,\hf,\hf),(-\hf,\hf,\hf,\hf)$, and $(\hf,-\hf,\hf,\hf)$.
The chirality is the product of the third and the fourth entries
and turns out to be the same as the hypermultiplet from the
untwisted sector.}

\subsection{Anomaly}

In 6D, there can exist a square anomaly. The anomaly of
$SO(2N)$ for $N\ge 5$ is\cite{frampton},
\begin{eqnarray}
 +2(N-4) & \mbox{ for the adjoint representation}, \nonumber \\
 -2^{N-5} & \mbox{ for the spinor representation},
\end{eqnarray}
with the normalization in units of the anomaly of
the left-handed vector representation. It is important that 
in this model we have gaugino and matter 
fermion with opposite chirality. In six dimension, Weyl spinor is 
self-dual, or charge conjugate of one has the same chirality of itself. 

We can check  the anomaly cancellation with the fermion spectrum
obtained in the preceeding section,

\begin{eqnarray} 
&{\cal A} = {\cal A}({\rm L-handed\ Adjoint})+{\cal A}({\rm 
R-handed\ spinor})\nonumber\\
&+ 16{\cal A}({\rm R-handed\ vector})= 8 + 8 +16(-1) = 0.
\nonumber
\end{eqnarray}

This again shows the consistency of having only one spinor
representation due to the nontrivial center of $SO(16)$ after
orbifolding with $Z_2$.

\subsection{$E_7\times SU(2)$}

In 6D the
square anomalies of {\bf 133}(adjoint) and {\bf 56} of $E_7$ are 
absent. As for the anomaly cancellation, therefore, 
$E_7$ in 6D is like $SO(N)$  in 4D. 
As we studied the $SO(16)$ model previous subsection, we can repeat a
similar analysis. The gauge group is $E_7\times SU(2)\times E_8'$
or $E_7\times SU(2)\times SO(16)'$.
The matter in the untwisted sector is listed in Table IV.
In this subsection, we will comment the $E_7\times SU(2)$ flavor
unification briefly. In the untwisted sector, matter fields of
Table IV arise. 
The states in Table IV constitute ({\bf 56,2}) of $E_7\times SU(2)$ 
since the
center of $E_7\times SU(2)$ is $Z_2\times Z_2$. We divided
the total number by 2, as we have done in the $SO(16)$ case.
The flavor 
group is $SU(2)$. Thus, we obtain the flavor doublet of {\bf 56}.
It is a 6D model. 
\begin{center}
Table IV. {\it Root vectors $p_I$ in untwisted sector transforming 
like $\alpha^1$.}
\vskip 0.3cm
\begin{tabular}{|cc|}
\hline
vector & number of states \\
\hline
$ (\underline{ 1~0}~\underline{\pm 1~0~0~0~0~0}) $ & (12,2) \\
$ (\underline{ -1~0}~\underline{\pm 1~0~0~0~0~0}) $ & (12,2) \\
$ ([\hf~\hf]~[\hf~\hf~\hf~\hf\hf\hf]) $ & (32,2) \\
\hline
\end{tabular}
\end{center}

In the twisted sector, we obtain following states satisfying
$(p+v)^2=3/2$,
\begin{eqnarray}
 &(\underline{ -1~0}~\underline{\pm 1~0~0~0~0~0}),\ \
  \mbox{ 2({\bf 12,1})} \nonumber\\
 &(-\hf~-\hf~[\hf~\hf~\hf~\hf\hf\hf]),\ \   \mbox{
(\bf{32,1})} \nonumber
\end{eqnarray}
which constitute ({\bf 56,1}). Since there are 16 fixed points,
there are 16 copies of {\bf (56,1)}.
If we somehow remove the {\bf 56}'s from the twisted sector, we
can have a four generation model. Indeed, we can devise such a
scheme by compactifying the 6th dimension by $S_1/Z_2$ or by the
Scherk-Schwarz mechanism. Since the number of components of spinors
in 6D and 5D are the same, by the Scherk-Schwarz mechanism for example, 
we just distinguish the representation property. We assign the 
anti periodic boundary condition for all the wave functions in 
the compactification of the 6th dimension.
The $2\pi$ rotation in the 6th dimension is embedded in the
$SU(2)$ group space so that the spinors of $SU(2)$ get an
extra minus sign for the $2\pi$ rotation. 
Therefore, sixteen ({\bf 56,1})'s are
projected out. In 5D we obtain an $E_7\times SU(2)$ model with
the matter ({\bf 56,2}).

The $E_7$ braching of {\bf 56} to $E_6$ representations is
\begin{equation}
{\bf 56}\rightarrow  {\bf 27}+\overline{\bf 27} 
+{\bf 1}+{\bf 1}.
\end{equation} 
Thus, it is possible to pick up two {\bf 27}'s in 4D by picking 
one {\bf 27} from the left-handed and one $\overline{\bf 27}$ 
from the right-handed. Then, we obtain a four generation model
since {\bf 56} is a flavor group doublet. 

\section{Conclusion}

We considered the $Z_2$ orbifold compactification of the
heterotic string to obtain a 6D $SO(16)$ model with one
spinor. We pointed out that the untwisted sector matter fields
should be properly counted by considering the center of
$SO(16)$. One $SO(16)$ spinor can split into two $SO(14)$
spinors, and a 5D $S_1/Z_2\times Z_2'$ compactification
leads to reasonable family unification models\cite{hk1}. 
We also commented on the $E_7\times SU(2)$ flavor unification.

\acknowledgments
We thank Paul Frampton and Kyuwan Hwang for useful
discussions. One of us(JEK) thanks Humboldt Foundation for the award. 
This work is supported in part by the BK21
program of Ministry of Education, and the KOSEF Sundo Grant.


\begin{thebibliography}{99}

\def\apj#1#2#3{Astrophys.\ J.\ {\bf #1} (#3) #2}
\def\ijmp#1#2#3{Int.\ J.\ Mod.\ Phys.\ {\bf #1} (#3) #2}
\def\mpl#1#2#3{Mod.\ Phys.\ Lett.\ {\bf A#1} (#3) #2 }
\def\nat#1#2#3{Nature\ {\bf #1} (#3) #2}
\def\npb#1#2#3{Nucl.\ Phys.\ {\bf B#1} (#3) #2}
\def\plb#1#2#3{Phys.\ Lett.\ {\bf B#1} (#3) #2}
\def\prd#1#2#3{Phys.\ Rev.\ {\bf D#1} (#3) #2}
\def\prl#1#2#3{Phys.\ Rev.\ Lett.\ {\bf #1} (#3) #2}
\def\prt#1#2#3{Phys.\ Rep.\ {\bf #1} (#3) #2}
\def\sjnp#1#2#3{Sov.\ J.\ Nucl.\ Phys.\ {\bf #1} (#3) #2}
\def\zp#1#2#3{Z.\ Phys.\ {\bf #1} (#3) #2}
\def\jhep#1#2#3{JHEP\ {\bf #1} (#3) #2}
\bibitem{georgi} H. Georgi, \npb{156}{126}{1979}.

\bibitem{kimsu7} J. E. Kim, \prl{45}{1916}{1980}; \prd{23}{2706}{1981}.

\bibitem{dixon} L. Dixon, J. Harvey, C. Vafa and E. Witten,
\npb{261}{1985}{651}; \npb{274}{285}{1986}; L. Ibanez, H. P.
Nilles, and F. Quevedo, \plb{187}{25}{1987}.

\bibitem{iknq} L. Ibanez, J. E. Kim, H. P. Nilles, and F. Quevedo,
\plb{191}{282}{1987}.

\bibitem{imnq} L. Ibanez, J. Mas, H. P. Nilles, and F.
Quevedo, \npb{301}{157}{1988}.

\bibitem{su51} S. M. Barr, \plb{112}{219}{1982}; J.-P. Derendinger,
J. E. Kim, and D. V. Nanopoulos, \plb{139}{170}{1984}.

\bibitem{ellis} I. Antoniadis, J. Ellis, J. S. Hagelin, and D. V.
Nanopoulos, \plb{231}{65}{1989}, and references cited 
therein.

\bibitem{barr} K. S. Babu, S. Barr, and B. Kyae, \prd{65}{115008}{2002}. 


\bibitem{hk1} K. Hwang and J. E. Kim, \plb{540}{289}{2002}.

\bibitem{candelas} P. Candelas, G. T. Horowitz, A. Strominger,
and E. Witten, \npb{258}{46}{1985}.

\bibitem{kawamura} Y. Kawamura, Prog. Theor. Phys. {\bf 103}
(2000) 613 [hep-ph/9902423].

\bibitem{bacry} H. Bacry, in {\it Lectures on Group Theory
and Particle Theory} (Gordon and Breach, New York, 1977) p.535.

\bibitem{shafi} G. Lazarides and Q. Shafi, \plb{115}{21}{1982}.

\bibitem{frampton} P. Frampton and T. Kephart, \prd{28}{1010}{1983}; 
S. Okubo and Y. Tosa, \prd{36}{2484}{1987}; A. Hebecker
and J. March-Russell, \npb{625}{128}{2002}.


\end{thebibliography}
\end{document}